\documentclass[twocolumn,trackchanges]{aastex7}
\usepackage{graphicx} 

\begin{document}
 
\title{Image Marker}
 
\author[orcid=0000-0001-5424-3698, gname=Ryan, sname=Walker]{Ryan Walker}
\altaffiliation{These two authors contributed equally to this work.}
\affiliation{Argonne National Laboratory, United States}
\email{astrorya@umich.edu}
 
\author[orcid=0009-0006-7664-877X, gname=Andi, sname=Kisare]{Andi Kisare}
\altaffiliation{These two authors contributed equally to this work.}
\affiliation{Argonne National Laboratory, United States}
\email{amkisare@gmail.com}
 
\author[orcid=0000-0001-7665-5079, gname=Lindsey, sname=Bleem]{Lindsey Bleem}
\affiliation{Argonne National Laboratory, United States}
\email{lbleem@anl.gov}
 
\begin{abstract}
    A wide range of scientific imaging datasets benefit from human inspection for purposes ranging from prosaic---such as fault identification and quality inspection---to profound, enabling the discovery of new phenomena. As such, these datasets come in a wide variety of forms, with diverse inspection needs. In this paper we present a software package, \texttt{Image Marker}, designed to help facilitate human categorization of images. The software allows for quick seeking through images and enables flexible marking and logging of up to nine different classes of features and their locations in files of FITS, PNG, TIFF, and JPEG format. Additional tools are provided to add text-based comments to the marking logs and for displaying external mark datasets on images during the classification process. As our primary use case will be the identification of features in astronomical survey data, \texttt{Image Marker} also utilizes standard World Coordinate Systems (WCS) headers embedded in FITS headers and TIFF metadata when available. The lightweight software, based on the Qt Framework to build the GUI application, enables efficient marking of thousands of images on personal-scale computers. We provide \texttt{Image Marker} as a Python package, and as Mac and Windows 11 executables. It is available \href{https://github.com/andikisare/imgmarker/}{on GitHub} or via pip installation.
\end{abstract}
 
 
\section{Statement of need\label{sec:intro}}
 
The rapid advancement in detector technology across all fields of science has led to larger and larger datasets without an equal increase in the number of scientists available to analyze the data. This imbalance of available work to available workers has led to a need for developing more efficient methods of parsing data. In response to large datasets in astronomy, projects like DES Exposure Checker \citep{DES_Exposure_Checker}, Space Warps \citep{Space_Warps}, and Galaxy Zoo \citep{Galaxy_Zoo} emerged to crowdsource classification and identification tasks in large datasets, the latter two using the Zooniverse framework \citep{Zooniverse}. Zooniverse offers the ability to easily outsource image identification and advanced classification statistics through the power of citizen science. This level of sophistication is not required, however, for projects which may involve fewer collaborators or for low-level data or algorithmic phases that are not suitable for a broader audience. Zooniverse also requires an internet connection. FitsMap \citep{FitsMap} takes a different approach with a focus on large images and their associated catalogs by hosting a web client on the user's local machine and displaying a reduced-scale image with catalog objects overlaid. While FitsMap has broad functionality, it does not contain a method for marking features in images and saving feature coordinates. Other software for viewing and analyzing data, like the widely-used SAO-DS9 \citep{SAO_DS9}, handle smaller datasets best.
 
\begin{figure*}[ht]
\epsscale{1.15}
\plotone{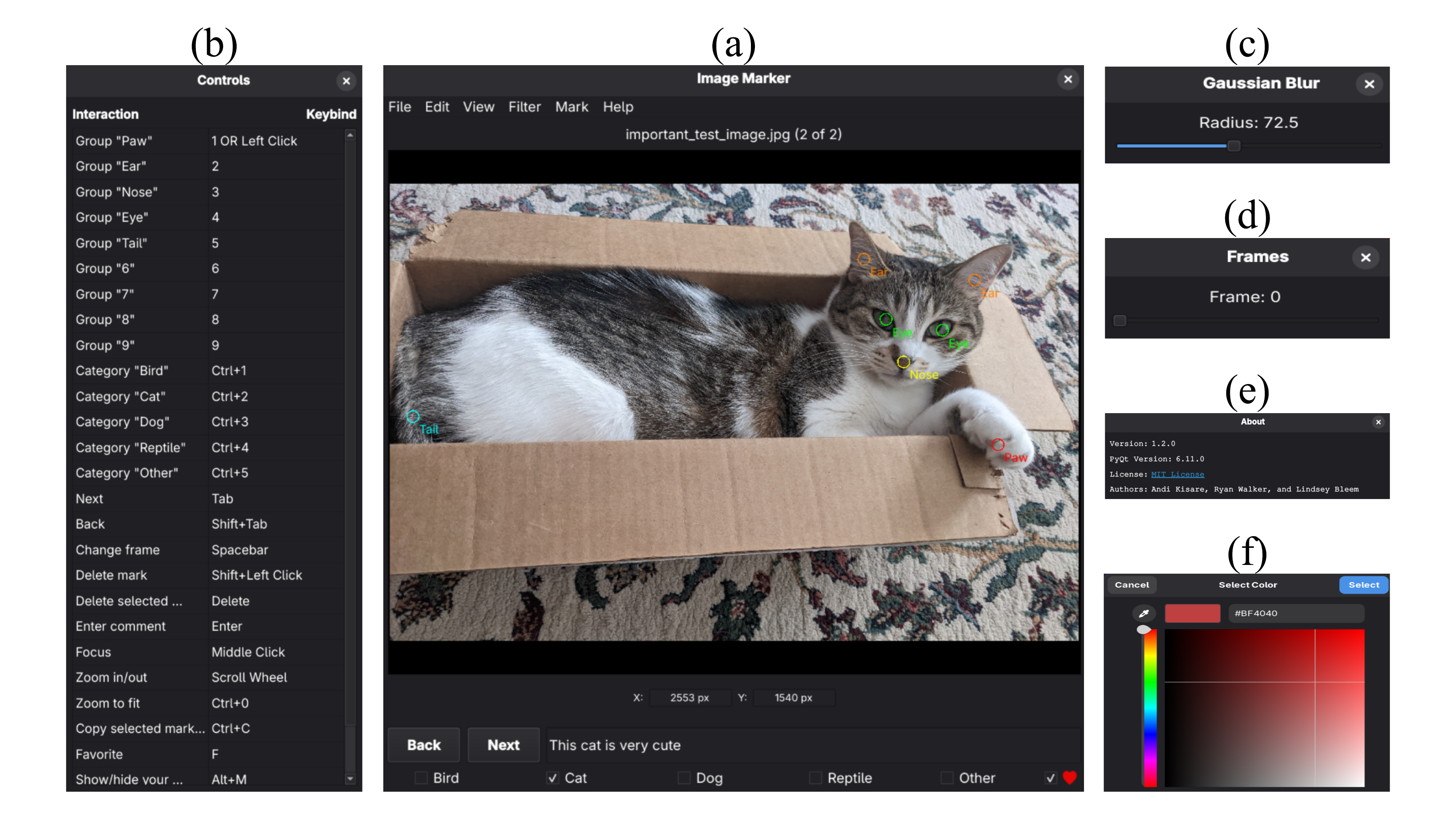}
\caption{Diagram of \texttt{Image Marker} windows highlighting main features of the application: (a) main window; nine different group marks are available for tagging features. Below the image users can read the pixel coordinates (and, if available, WCS coordinates) of the cursor. Note that a comment has been written in the main comment box in the center. (b) controls window; updates group and category names when they are customized, helping keep track of what buttons are for which group or category. Other shortcuts are shown as well. (c) Gaussian blur window; note that blur has not been applied to the example image in (a). (d) frames window; for selecting frames in multi-extension images/FITS files. (e) about window; displays basic information about the user's installation of \texttt{Image Marker}. (f) color picker window; used to select the color of an imported \textit{catalog} of \textit{marks}. Window themes are dependent on the user's operating system.
\label{fig:figure1}
}
\end{figure*}
 
\texttt{Image Marker} is a tool specifically designed for quickly scanning images and tagging locations in the images. It is run locally, has a user-friendly interface, a fail-safe saving mechanism, and also includes a suite of features and options for customization, image manipulation, and testing user consistency (see Figure \ref{fig:figure1}) for application interface). By sharing an \texttt{Image Marker} configuration file and data with other users, \texttt{Image Marker} also allows joint analyses of datasets at the expense of requiring manual sharing of (small) files after marking is completed.
These output files are easily aggregated and analyzed using standard coding or spreadsheet tools. Local analysis enables trained observers to quickly scan through images with loading times not limited by internet connections and thus enables faster identifications or classifications.
 
\subsection{Our use case\label{subsec:ouruses}}
\begin{figure*}[ht]
\epsscale{1.15}
\plotone{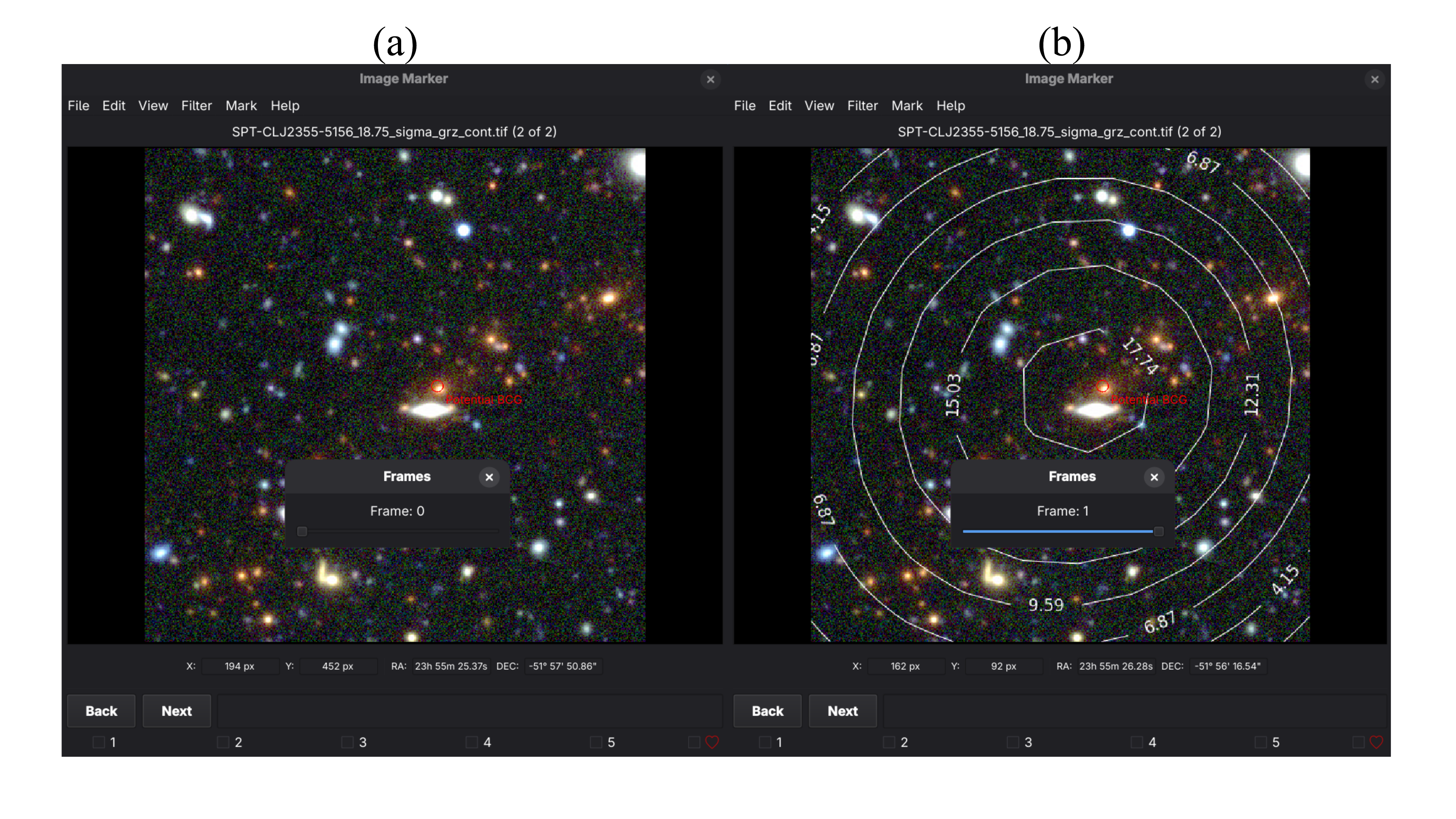}
\caption{Optical \textit{grz} band image of a galaxy cluster from the SPT-3G survey \citep[Optical images from DeCALS][]{Dey19}. (a, Left) We display the first frame of the image file with just optical image data. (b, Right) The second frame of the image file, which contains the optical image data with contours overlaid indicating the SZ detection signal-to-noise from SPT-3G. The human-selected candidate BCG is denoted by the red mark in both images.
\label{fig:figure2}
}
\end{figure*}

Galaxy clusters are useful probes for peering into the early Universe. One method of connecting cluster observables to theoretical models requires selecting a cluster center \citep[using e.g., weak gravitational lensing; see reviews in][]{Allen11, Umetsu20}, the most common choice being to choose the Brightest Cluster Galaxy (BCG), namely the brightest or largest, centralized galaxy in the cluster. However, automatic BCG selection algorithms typically fail 10-20\% of the time, so human inspection plays an important role in both validating these algorithms and improving centering choices when they fail \citep{Rozo14, Ding-miscentering, Kelly-miscentering}. The SPT-3G camera has surveyed $\sim$10,000 square degrees of the Southern sky \citep{Sobrin22, Prabhu24} at millimeter-wavelengths, giving plenty of data to inspect. Our first use case for \texttt{Image Marker} is the identification of BCGs in the SPT-3G cluster sample (see \autoref{fig:figure2}) using optical image data from DeCALS \citep{Dey19}. The validated BCG dataset from thousands of clusters \citep[$>$7,100 galaxy clusters expected in the SPT-3G main survey alone][]{Bleem26} also offers the opportunity to develop and test machine learning BCG identification tools that can be applied not only to the SPT-3G sample but also cluster samples from other surveys \citep[e.g.,][]{Bulbul24, Hilton21, LSST}.
 
\subsubsection{Broader use cases\label{subsubsec:broaderuses}}
 
While \texttt{Image Marker} was initially designed with the above use cases in mind, we have found it valuable as a general tool for inspecting data products and validating algorithmic development. As a second example usage, the ability to rapidly scan hundreds of small thumbnail cutouts in a matter of minutes, mark problematic locations, and easily read in lists of these locations, helped us to improve data cleaning for an upcoming analysis of SPT-3G data in the Euclid Deep Field South region. This broad applicability motivated us to publicly release the software.
 
\begin{acknowledgements}
We thank Keren Sharon, Mike Gladders, and Giulia Campitiello for helpful suggestions on features to include in \texttt{Image Marker}, Will Hicks for help with compiling \texttt{Image Marker} on an Intel based Mac, and Florian Keruzore and the JOSS referees, Erfan Nourbakhsh and Alex Lyttle, for providing helpful comments.
This work was supported in part by the U.S. Department of Energy, Office of Science, Office of Workforce Development for Teachers and Scientists (WDTS) under the Science Undergraduate Laboratory Internships (SULI) program. Work at Argonne National Lab is supported by UChicago Argonne LLC, Operator of Argonne National Laboratory (Argonne). Argonne, a U.S. Department of Energy Office of Science Laboratory, is operated under contract no. DE-AC02-06CH11357.
We thank the SPT-3G collaboration for the use of the Sunyaev Zel'dovich detection contours displayed on the image in Figure \ref{fig:figure2}).
This work made use of \href{http://www.astropy.org}{Astropy}: a community-developed core Python package and an ecosystem of tools and resources for astronomy \citep{astropy13, astropy18, astropy22}; Pillow \citep{clark15}; SciPy \citep{Virtanen20}; NumPy \citep{Harris20}; and PyQt \citep{PyQt}.
\end{acknowledgements}
 
\bibliography{main}
\bibliographystyle{aasjournalv7}
 
\end{document}